# Magnetic Moment of the Proton

**G. González-Martín*,  I.Taboada**

Departamento de Física, Universidad Simón Bolívar, Apartado 89000, Caracas 1080-A, Venezuela.

**and J. González**

Physics Department, Northeatern University, Boston, U.S.A.

*Webpage: http://prof.usb.ve/ggonzalm/

The magnetic moment of the proton is calculated using a geometric unified theory. The geometry determines a generalized Pauli equation showing anomalous magnetic terms due to the triplet proton structure. The theoretical result gives a bare anomalous Landé gyromagnetic g-factor close to the experimental value. The necessary radiative corrections should be included in the actual theoretical dressed value. The first order correction raises the value to 2(2.7796). Similarly we obtain for the neutron gyromagnetic g-factor the value 2(1.9267).

# 1. Introduction.

The proton triple structure may be related to a non-linear geometric theory that unifies gravitation and electromagnetism that offers the possibility of representing other interactions by a sector of the connection [1,2,3]. It has been shown that new electromagnetic consequences of the theory lead to quanta of electric charge and magnetic flux, providing a plausible explanation to the fractional quantum Hall effect [4,5]. On the other hand we also have considered an approximation to this geometric non linear theory [6] where the microscopic physical objects (geometric particles) are realized as linear geometric excitations, geometrically described in a jet bundle formalism shown to lead to the standard quantum field theory techniques. These geometric excitations are essentially perturbations around a non-linear geometric background space solution, where the excitations may be considered to evolve with time. In this framework, a geometric particle is acted upon by the background connection and is never really free except in absolute empty background space (zero background curvature). The background space carries the universal inertial properties which should be consistent with the ideas of Mach [7] and Einstein [8] that assign fundamental importance of far-away matter in determining the inertial properties of local matter including the inertial mass. We may interpret the geometric excitations as geometric particles and the background as the particle vacuum. This is a generalization of what is normally done in quantum field theory when particles are interpreted as vacuum excitations. The vacuum is replaced by a background solution that we call the substratum. The field equation of the theory admits a constant connection and current substratum solution determining a geometric symmetric curved space [9,10].

The geometry is related to a connection $\Gamma$ in a principal fiber bundle $(E,M,G)$. The structure group $G$ is SL(4,R) and the even subgroup $G_+$ is $SL_1(2,C)$. The subgroup $L$ (Lorentz) is the subgroup of $G_+$ with real determinant, in other words, SL(2,C). There is only another subgroup $P$ in the possible group chains $G\supset H\supset L$, which is Sp(4,R). The holonomy groups of the connection may be used geometrically to classify the interactions contained in the theory. The subgroup chain $SL(4,R) \supset Sp(4,R) \supset SL(2,C)$ characterizes a chain of subinteractions with reducing algebraic sectors. Electromagnetism is associated to an $SU(2)_Q$ subgroup. Matter is represented by a current $J$ expresed in terms of a principal fiber bundle section $e$, which is related by charts (coordinates) to matrices of the group SL(4,R). This structure defines a frame of SL(4,R) spinors over space-time. The connection is an sl(4,R) 1-form that acts naturally on the frame $e$ (sections).

There are certain important problems in the theory that should be discussed, as follows: 1- To find the role played by the extra generators or connection components in the theory as non classical interactions, possibly nuclear interactions. In particular, since only one of the three $SU(2)_Q$ generators is observable as a long range electromagnetic potential 4-vector $A$, the significance of the other 2 unobservable short range connection components should be clarified; 2- To further discuss the equations of motion for a complete $G$-system which should display additional effects in comparison with the simpler equations for a $P$-system.

Here we address these problems. When we write the equations of motion for a complete $G$-system, we find that we are able to calculate the bare magnetic moment of the fundamental (hadron) particle associated to the strong nuclear interactions, in a non relativistic approximation.

# 2. General equations of Motion.

The non linear equations for the connection have a solution that we have called substratum [9, 10] which may be expressed, in a particular reference system, as a tensor valued form $\Lambda$ in terms of a constant solution $m_0$

$$\Lambda = \frac{-m_0}{4}J \ . \tag{1}$$

Using the definition of mass from the energy $J.\Gamma$ [11] the calculated mass corresponding to this substratum solution is

$$m = \tfrac{1}{4}\text{tr}\left(J^\mu \Gamma_\mu\right) = \tfrac{1}{4}\text{tr}\left(J^\mu \Lambda_\mu\right) = \tfrac{1}{4}\text{tr}\left(\left(e^{-1}\kappa^\mu e\right)\left(\frac{-m_0}{4}e^{-1}\kappa_\mu e\right)\right) = \tfrac{1}{4}\text{tr}\left(m_0 I\right) = m_0 \ . \tag{2}$$

We may define a second connection by subtracting the tensor valued form $\Lambda$ from the geometric connection

$$\hat{\Gamma} \equiv \Gamma - \Lambda = \Gamma - \left(\frac{-m}{4}J\right). \qquad (3)$$

Using the new connection the equation of motion explicitly displays a substratum mass term

$$\kappa^\mu \nabla_\mu e = \kappa^\mu \left(\partial_\mu - e\Gamma_\mu\right) = \kappa^\mu \left(\partial_\mu - e\hat{\Gamma}_\mu - e\left(\frac{-m}{4}J_\mu\right)\right) = \kappa^\mu \hat{\nabla}_\mu e - me = 0 \qquad (4)$$

and the equations for the even and odd parts of a G-system are, using the corresponding expressions in the appendix,

$$\sigma^\mu \nabla_\mu \xi = \sigma^\mu \bar{\eta}^\dagger \hat{\Gamma}_{-\mu} - m\eta \ , \qquad (5)$$

$$-\bar{\sigma}^\mu \nabla_\mu \eta = \bar{\sigma}^\mu \bar{\xi}^\dagger \hat{\Gamma}_{-\mu} - m\xi \ . \qquad (6)$$

We now designate different vector potentials in the connection with the following notation: $^eA$ is the electromagnetic potential corresponding to the even part of the SU(2)$_Q$ sector of the connection; $^oA$ is the complementary odd part of the SU(2)$_Q$ sector; $\Gamma$ is the SL(2,C) sector of the connection which determines an $L$ covariant derivative; $\Upsilon$ is the complementary part. Both $^oA$ and $\Upsilon$ are complex elements with real part along $\kappa^0$ and imaginary part along $\kappa^0\kappa^5$. The first equation may be written as

$$\sigma^\mu \left(\partial_\mu \xi - i\,^eA_\mu \xi - \xi \Gamma_\mu \right) = \sigma^\mu \,^oA_\mu \bar{\eta}^\dagger + \sigma^\mu \bar{\eta}^\dagger \Upsilon_\mu - m\eta \ . \qquad (7)$$

Inserting the unit $-i^2$ in the left side of this equation

$$-\sigma^\mu \left(i\partial_\mu (i\xi) + \,^eA_\mu (i\xi) - (i\xi)(i\Gamma_\mu)\right) = \sigma^\mu \,^oA_\mu \bar{\eta}^\dagger + \sigma^\mu \bar{\eta}^\dagger \Upsilon_\mu - m\eta \qquad (8)$$

and using

$$\bar{\eta} = -\eta \ , \qquad (9)$$

$$\bar{\xi} = -\xi \ , \qquad (10)$$

we get

$$\sigma^\mu \left(i\nabla_\mu + \,^eA_\mu\right)(i\xi) = \sigma^\mu \,^oA_\mu \eta^\dagger + \sigma^\mu \eta^\dagger \Upsilon_\mu + m\eta \ . \qquad (11)$$

Similarly for the second equation we get after multiplying by $i$

$$-\bar{\sigma}^\mu \left(i\partial_\mu \eta + \,^eA_\mu \eta - \eta(i\Gamma_\mu)\right) = \bar{\sigma}^\mu \,^oA_\mu \left(i\bar{\xi}^\dagger\right) + \bar{\sigma}^\mu \left(i\bar{\xi}^\dagger\right)\Upsilon_\mu - m(i\xi) \ , \qquad (12)$$

$$\bar{\sigma}^\mu \left(i\nabla_\mu + \,^eA_\mu\right)(\eta) = \bar{\sigma}^\mu \,^oA_\mu \left(i\xi^\dagger\right) + \bar{\sigma}^\mu \left(i\xi^\dagger\right)\Upsilon_\mu + m(i\xi) \ . \qquad (13)$$

If we define

$$\varphi \equiv \tfrac{1}{\sqrt{2}}(\eta + i\xi) \ , \qquad (14)$$

$$\chi \equiv \tfrac{1}{\sqrt{2}}(\eta - i\xi) \ , \qquad (15)$$

we can write the equations in the following form

$$\left(i\nabla_0 + \,^eA_0\right)\varphi - \sigma^m \left(i\nabla_m + \,^eA_m\right)\chi = \,^oA_0 \chi^\dagger + \sigma^m \,^oA_m \varphi^\dagger + $$
$$+ \chi^\dagger \Upsilon_0 + \sigma^m \varphi^\dagger \Upsilon_m + m\varphi \ , \qquad (16)$$

$$(i\nabla_0 + {}^eA_0)\chi - \sigma^m(i\nabla_m + {}^eA_m)\varphi = -{}^oA_0\varphi^\dagger - \sigma^m{}^oA_m\chi^\dagger +$$
$$-\varphi^\dagger\Upsilon_0 - \sigma^m\chi^\dagger\Upsilon_m - m\chi \quad . \tag{17}$$

## 3. Non Relativistic Approximation

We now take the equations as linearized equations around the substratum solution with a connection fluctuation representing self interactions. For a non relativistic approximation, the low velocities and corresponding boosts are of order $v/c$. We may neglect the terms of order $v/c$, which are the boost $\Upsilon$ and the hermitian parts of $\eta$ and $\xi$ leaving their antihermitian parts

$$\eta^\dagger = -\eta \quad , \tag{18}$$

$$\xi^\dagger = -\xi \quad , \tag{19}$$

and we obtain

$$\chi^\dagger = \tfrac{1}{\sqrt{2}}(\eta^\dagger + i\xi^\dagger) = \tfrac{1}{\sqrt{2}}(-\eta - i\xi) = -\varphi \quad , \tag{20}$$

$$\varphi^\dagger = \tfrac{1}{\sqrt{2}}(\eta^\dagger - i\xi^\dagger) = \tfrac{1}{\sqrt{2}}(-\eta + i\xi) = -\chi \quad . \tag{21}$$

As usually done in relativistic quantum mechanics [12] in a non relativistic approximation, we let

$$\varphi \to \varphi e^{-imt} \quad , \tag{22}$$

$$\chi \to \chi e^{-imt} \quad , \tag{23}$$

obtaining slowly varying functions of time with equations

$$(i\nabla_0 + {}^eA_0)\varphi - \sigma^m(i\nabla_m + {}^eA_m)\chi = -{}^oA_0\varphi - \sigma^m{}^oA_m\chi +$$
$$-\varphi\Upsilon_0 - \sigma^m\chi\Upsilon_m \quad , \tag{24}$$

$$(i\nabla_0 + {}^eA_0)\chi - \sigma^m(i\nabla_m + {}^eA_m)\varphi = {}^oA_0\chi + \sigma^m{}^oA_m\varphi +$$
$$+\chi\Upsilon_0 + \sigma^m\varphi\Upsilon_m - 2m\chi \quad . \tag{25}$$

The equations become after recognizing the space component ${}^oA_m$ as an odd electromagnetic vector potential.

$$(i\nabla_0 + {}^eA_0 + {}^oA_0)\varphi - \sigma^m(i\nabla_m + {}^eA_m - {}^oA_m)\chi = 0 \quad , \tag{26}$$

$$(i\nabla_0 + {}^eA_0)\chi - \sigma^m(i\nabla_m + {}^eA_m + {}^oA_m)\varphi = -2m\chi + {}^oA_0\chi \quad , \tag{27}$$

where $\chi$ is the small component relative to the large component $\varphi$. We neglect the small terms, as usual the $\chi$ terms in the last equation unless multiplied by $m$, and substitute the expression for $\chi$ in equation (26). The result is

$$(i\nabla_0 + {}^eA_0 + {}^oA_0)\varphi -$$
$$\frac{\sigma^m(i\nabla_m + {}^eA_m - {}^oA_m)\sigma^n(i\nabla_n + {}^eA_n + {}^oA_n)\varphi}{2m} = 0 \quad . \tag{28}$$

Since we have the well known relation

$$\sigma.a\,\sigma.b = a.b + i\sigma.(a\times b) \quad , \tag{29}$$

substitution in equation (28) gives

$$i\nabla_0 \varphi = \left[ \frac{\left(i\nabla + {}^eA - {}^oA\right) \cdot \left(i\nabla + {}^eA + {}^oA\right)}{2m} - \left({}^eA_0 + {}^oA_0\right) + \right. $$
$$\left. -\frac{\sigma \cdot \left(\nabla \times \left({}^eA + {}^oA\right)\right)}{2m} + \frac{\sigma \cdot \left({}^oA \times \nabla\right)}{m} + \frac{i {}^oA \times {}^eA}{m} \right] \varphi \quad . \tag{30}$$

which is Pauli's equation [13] with extra terms depending on the even and odd vector potentials ${}^eA$, ${}^oA$.

## 4. Magnetic Moment Term

Equation (30) displays the remarkable geometrical structure of triplets. According to physical geometry [14] a $G$-solution should include the three SU(2)$_Q$ generators. We may associate the effect of a combination of three SU(2)$_Q$ connection components, one for each possible $P$ in $G$, as three classically equivalent electromagnetic potentials $A$'s [2, 3]. It has been shown [14], in this geometry, that all long range fields correspond to fields associated to the fiber bundle obtained by contracting the structure group SL(4,R) to its even subgroup SL$_1$(2,C) which in turn correspond to classical fields. Therefore the long range component of the sl(4,R)-connection coincides with the long range component of an sl$_1$(2,C)-connection corresponding to an even subalgebra sl(2,C)⊕u(1) of sl(4,R) [14] related to gravitational and electromagnetic fields. In fact any direction in the tridimensional geometric electromagnetic su(2)$_Q$ subalgebra may be identified as a valid direction corresponding to this remaining long range classical electromagnetic u(1). There is no preferred direction in su(2)$_Q$. If we observe a long range electromagnetic field, we may always align the classical field $A$ with any of the 3 geometric electromagnetic $\kappa$ generators in su(2)$_Q$, or a linear combination, by performing SU(2)$_Q$ transformation. Nevertheless the two extra $A$'s should make additional contributions to the magnetic energy of the short range $G$-system, as shown in equation (30), and therefore to its corresponding magnetic moment.

The electromagnetic subgroup is SU(2)$_Q$, similar to the spin subgroup SU(2)$_S$. The group itself, as a fiber bundle (SU(2),S$^2$,U(1)) carries its own representations. The base space is the coset SU(2)/U(1) which is the bidimensional sphere S$^2$. The fiber is an arbitrary even subgroup U(1). The action of this electromagnetic SU(2)$_Q$ is a multiplication on the fiber by an element of the U(1) subgroup and a translation on the base space S$^2$ by the action of the SU(2)$_Q$ group Casimir operator, representing a squared total SU(2)$_Q$ rotation.

This action is not as simple as translations in flat spaces, but rather has complications similar to those associated with angular momentum due to the SU(2) group geometry. The orientation of directions in SU(2) is quantized. In particular only one component of the electromagnetic rotation generator $E$ commutes with the group Casimir operator $E^2$. This operator acts on the symmetric coset S$^2$ becoming the Laplace-Beltrami operator on the coset. Its eigenvalues are geometrically related to translations on S$^2$ in the same manner as the eigenvalues of the usual Laplace operator are related to translations on the plane. There are definite simultaneous eigenvalues, with the Casimir operator, only along any arbitrary *single* su(2) direction, which we have taken as the direction of the even generator ${}^eE$. The generator $E$ may be decomposed in terms of the complementary odd component ${}^oE$. We can not decompose the $A$ connection into components with definite expectation values. The splitting of $A$ into an even part ${}^eA$ and an odd part ${}^oA$ represents, respectively, the splitting of the group action into its vertical even action on the fiber and a complementary odd translation on the base S$^2$.

Consider that the exponential functions $e^{k \cdot r}$ form a representation of the translation group on the plane. The magnitude of the translation $k$ is determined by the eigenvalue of the Laplace operator $\triangle$,

$$\Delta e^{k \cdot x} = \lambda e^{k \cdot x} = k^2 e^{k \cdot x} \quad , \tag{31}$$

where the absolute value of $k$ is

$$|k| = \left(\delta^{mn} k_m k_n\right)^{1/2} \quad . \tag{32}$$

The odd subspace of su(2)$_Q$, spanned by the two compact odd electromagnetic generators in ${}^oE$, is isomorphic to the odd subspace of the quaternion algebra, spanned by its orthonormal set $q^a$. Associated to this orthonormal set we have the Dirac operator $q \cdot \nabla$ on a curved bidimensional space. This Dirac operator is the representation of ${}^oE$ on the vector functions on the sphere. We obtain for this action of ${}^oE$, if we separate the wave function $\varphi$ into the SU(2)$_S$ eigenvector $\phi$ and the SU(2)$_Q$ eigenvector $\psi$, and use the fact that the Levi-Civita connection is symmetric,

$$-\left(q^a \nabla_a\right)^2 \psi = q^a q^b \nabla_a \nabla_b \psi = q^{(a} q^{b)} \nabla_a \nabla_b \psi + q^{[a} q^{b]} \nabla_a \nabla_b \psi$$
$$= g^{ab} \nabla_a \nabla_b \psi = \Delta \psi = i(i+1)\psi \quad . \tag{33}$$

This equation shows that the squared curved Dirac operator is the Laplace-Beltrami or Casimir operator, with eigenvalues equal to the squared translation magnitude on the odd sphere S². Therefore the odd electromagnetic generator $^oE$ is related to translations on the electromagnetic S². In general quaternion $^oE$ squared is the Casimir quaternion $C^2$ which acts on the coset SU(2)/U(1). The $^oE$ direction in the odd tangent plane is indeterminable because there are no odd eigenvectors common with $^eE$ and $E^2$ and therefore no definite values for the generators in this plane. Nevertheless the absolute value of this quaternion must be the square root of the absolute value of the Casimir quaternion, which also corresponds to the squared total rotation $E^2$. In the orthonormal base of the common eigenvectors of $^eE$ and $E^2$, the absolute values of quaternions $^eE$ and $E^2$ are the respective eigenvalues. The absolute values of $^eE$ and $^oE$ define a polar angle $\theta$ in the su(2) algebra. We conclude that the electromagnetic generator has an indefinite azimuthal direction but a quantized polar direction determined by the possible translation values. Therefore we obtain

$$\frac{|^oE|}{|^eE|} = \frac{|E^2|^{1/2}}{|^eE|} = \frac{\sqrt{i(i+1)}}{n} \equiv \tan\theta \quad . \tag{34}$$

The internal direction of the potential $A$ must be along the possible directions of the electromagnetic generator $E$ in $su(2)_Q$. The $A$ components must be proportional to the possible even and odd translations. In consequence the total $A$ vector must lie in a cone defined by a quantized polar angle $\theta$ relative to an axis in the even direction and an arbitrary azimuthal angle. The energy term in equation (30) becomes

$$U = -\frac{\sigma}{2m}\cdot\left(\nabla\times\left(^eA + {}^oA\right)\right)\varphi = -\left(1 + \frac{(i(i+1))^{1/2}}{n}\right)\frac{\sigma}{2m}\cdot\left(\nabla\times {}^eA\right)\varphi \quad . \tag{35}$$

The fundamental state representing a proton is the $SU(2)_Q$ state with charge +1, corresponding to electromagnetic rotation eigenvalues ½, ½. In terms of the even magnetic field the energy becomes,

$$U = -\left(1 + \frac{(½(½+1))^{1/2}}{½}\right)\frac{\sigma\cdot {}^eB}{2m}\varphi = -\frac{(1+\sqrt{3})}{2m}\sigma\cdot {}^eB\varphi \quad . \tag{36}$$

Associated to the second √3 term in the parenthesis we have introduced the effective direction angle $\theta$ of the total generator, for a $G$-system, in the su(2) algebra. The value of $\theta$ is $\pi/3$. Statistically this direction corresponds to the expected (mean) value of the projected component, of a random *classical* direction, along the chosen even direction.

The first term in the parenthesis is related to the $P$-system which has only a U(1) electromagnetic subgroup and the complications due to SU(2) are not present. The complementary odd subspace corresponding to S² does no exist. The orientation of the complete electromagnetic connection $A$ can always be taken along the definite even direction defined by the physical u(1) algebra. The angle $\theta$ may be taken equal to zero. It corresponds to a $P$-system, associated to the electron, with only a $\kappa^0$ electromagnetic component. In this case the energy reduces to

$$U = \frac{-1}{2m}\sigma\cdot\left(\nabla\times {}^eA\right) = \frac{-1}{2m}\sigma\cdot {}^eB \quad . \tag{37}$$

If it were possible to make a transformation that aligns the internal direction $\theta$ along the even direction everywhere, we really would be dealing with a $P$-system because we actually would have restricted the connection to a $P$ subgroup. A $P$-system provides a preferred direction in the subgroup $SU(2)_Q$ of $G$, the only electromagnetic generator $\kappa^0$ of the associated $P$ group. If we use a $P$-test-particle (an electron) to interact with an external magnetic field $B$, we actually align the long range component $A$ with this preferred direction. In this manner we may explain the long-range physical experiments using an abelian electromagnetic field equation with a remote source and a Dirac equation. If we use a $G$-test-particle (a proton) to interact with an external field, the most we can do is to align the classical long-range component $A$ with a random direction in su(2). Part of the total internal field $^iB$ is only observable in a short range region. To each *B* direction related by an SU(2)$_Q$ transformation, there corresponds a "*partner*" spin direction, defining associated scalar products $\sigma \cdot B$. The additional scalar products, relative to a *P*-system, arise from the two additional geometric fields, and/or, equivalently, from the two additional copies of Pauli matrices $\kappa^n \kappa$ in the universal Clifford algebra, determined by the additional spin-magnetic directions $\kappa$, which originally were along $\kappa^1 \kappa^2 \kappa^3$ and $\kappa^5$. The additional spin operators, introduced by the short-range "*non-classical*" electromagnetic potentials, represent extra internal current loops that generate an "anomalous" increase of the intrinsic magnetic energy.

We have calculated the magnetic energy of a free *G*-system, with zero external field *B*, in terms of the observable even internal field $^eB$ whose internal direction coincides with the eventual internal direction of an external field *B*. In this gedenken experiment, the magnetic moment *is defined* as the partial derivative of the magnetic energy, produced by the total internal generalized magnetic field $^iB$, with respect to the even component $^eB$, in whose internal direction would align the internal direction of the external *B* field,

$$\mu^n \equiv -\frac{\partial U}{\partial\, ^eB_n} \quad . \tag{38}$$

The standard physical methods of magnetic moment measurement are nuclear paramagnetic resonance [15, 16], molecular beams and optical spectroscopy [17]. In a real physical experiment, when the sample is placed in the external field, there is a change in the magnetic energy of either a *G*-system or a *P*-system. For both systems, our test particle will respond to an external field, sensing a variation of the electromagnetic field linked to the test particle, in the internal direction of the external magnetic *B* field which coincides with the internal direction of the even $^eB$ field. The change in the magnetic energy, after restoring in the equation the fundamental physical constants *e*, $\hbar$, *c* which are all equal to 1 in our geometric units, is

$$\Delta U = \frac{dU}{dB}\Delta B = \frac{\partial U}{\partial\, ^eB}\Delta B = -\frac{(1+\tan\theta_t)e\hbar}{2m_ic}\sigma\cdot B$$
$$= -2(1+\tan\theta_t)\left(\frac{e\hbar}{2m_ic}\right)\frac{S\cdot B}{\hbar} = -g_i\mu_i\frac{S\cdot B}{\hbar} \tag{39}$$

where the variation seen by the test-particle is equal to the external field *B*, the internal angle $\theta_t$ is zero for the electron or $\pi/3$ for the proton and $m_i$ is the respective mass. This expression defines the magneton (atomic or nuclear) $\mu_i$ and the anomalous Landé gyromagnetic factor $g_i$.

The bare magnetic moment is given in general by

$$\mu^n = 2(1+\tan\theta_t)\left(\frac{e\hbar}{2m_ic}\right)\frac{\sigma^n}{2} = g_i\mu_i\frac{S^n}{\hbar} \tag{40}$$

in terms of the charge, mass and angle corresponding to the proton or electron. For the proton we obtain

$$\mu^n = 2(1+\sqrt{3})\left(\frac{e\hbar}{2m_pc}\right)\frac{\sigma^n}{2} = g_p\mu_N\frac{S^n}{\hbar} \tag{41}$$

determining the proton anomalous gyromagnetic factor (2.732) and the nuclear magneton. For the electron we obtain

$$\mu^n = -2\left(\frac{e\hbar}{2m_ec}\right)\frac{\sigma^n}{2} = g_e\mu_B\frac{S^n}{\hbar} \tag{42}$$

determining the electron anomalous gyromagnetic factor (-2) and the Bohr magneton. These identifications give the bare magnetic moments, from the magnetic energy in the geometric generalized Dirac equation, for the only two stable charged fundamental fermionic excitations in the geometric theory, the electron and the proton. They require radiative corrections to obtain the experimental values.

In QED the calculated values of the zero order magnetic moment for the electron and the proton, given by their external Coulomb field scattering diagrams, are singular. For both cases, the interaction hamiltonian for the external Coulomb field differs by the additional anomalous coefficient *g*, determined by equation (40). The effect of this difference is to adjoin the coefficient *g* to the external vertex in the diagram. Thus after renormalization the zero order magnetic moment values for the proton and electron are proportional by the factor of order zero $g_0$. The radiative corrections for the electron have been calculated by Schwinger [18, 19, 20]. For both, electron and proton, the first order corrections determined exclusively by the vertex part of the Coulomb scattering diagram are proportional to the corresponding zero order terms. For the electron the vertex correction diagram, formed by an internal photon line between the two internal fermion lines, gives Schwinger's correction factor $\alpha/2\pi$. For the proton the triple electromagnetic U(1) structure present in the SU(2) sector of the geometric interaction operator $J\bullet\Gamma$, determines three standard $j.A$ coupling terms. The triple structure is also present in the noncompact sector of the algebra [21]. This indicates that a full proton excitation description requires three boost momenta $k_i$. Consequently a full description of the external Coulomb field scattering diagram requires a pair of internal fermion boost momentum triplets, instead of simply the pair of electron momenta. There are additional U(1) radiative processes between the *six* internal fermion boost lines corresponding to the two triplets. Therefore, there are multiple additional internal vertex diagrams obtained by permutation of the internal photon line among the six necessary internal fermion lines, all contributing to the first order correction. The multiplicity of the vertex corrections of the *6* fermion boost lines taken *2* at a time is

$$M = \frac{n!}{p!(n-p)!} = \frac{6!}{2!4!} = 15 \quad . \tag{43}$$

There should be *15* vertex corrections, each equal to Schwinger's value, $\alpha/2\pi$, because of the SU(2) group equivalence. The total first order radiative contribution to the proton magnetic moment term gives

$$\frac{g_1}{2} = \frac{g_0}{2}\left(1 + \frac{15\alpha}{2\pi}\right) = 2.7321(1+0.01742) = 2.7796 \tag{44}$$

which reduces to *0.5%* the discrepancy with the experimental value, equal to *2.7928*.

In a similar manner we may calculate the geometric magnetic moment of a combination of *G*, *P* and *L* excitations. The total electromagnetic potential field of this excitation system is the sum of *A*, the SU(2)$_Q$ potential of the *G*-excitation, and $A_U$, the ***different*** U(1) potential of the *P*-excitation. Let $\varphi$ be the common frame (½) eigenfunction of the potentials $A$ and $A_U$. Thus, we may write the expression for the total magnetic energy, equation (35), in the following form,

$$U = -\frac{\sigma}{2m}\bullet\left(\nabla\times\left(A_U + {}^eA + {}^aA\right)\right)\varphi = -\left(1 + \frac{\left(i(i+1)\right)^{1/2}}{n+n_U}\right)\frac{\sigma}{2m}\bullet\left(\nabla\times\left(A_U + {}^eA\right)\right)\varphi \quad , \tag{45}$$

which becomes, in terms of the total even magnetic field ${}^eB$,

$$U = -\left(1 + \left(\tfrac{1}{2}(\tfrac{1}{2}+1)\right)^{1/2}\right)\frac{\sigma\bullet{}^eB}{2m}\varphi = -\frac{\left(1+\sqrt{3}/2\right)}{2m}\sigma\bullet{}^eB\varphi \quad . \tag{46}$$

The mass *m* corresponds to a *G*-excitation mass or proton mass $m_p$. We assume the resultant charge distribution preserves the dominant magnetic effect of the less massive *E*-excitation (electron). The corresponding expression for the magnetic moment, in terms of the nuclear magneton, is

$$\mu^n = -2\left(1+\sqrt{3}/2\right)\left(\frac{e\hbar}{2m_p c}\right)\frac{\sigma^n}{2} = g_n\mu_N \frac{S^n}{\hbar} \quad . \tag{47}$$

The radiative correction of this value should account for the presence of the additional *P*-excitation in the system. This indicates that the excitation description requires four electronic boost momenta instead of the three momenta required for the proton excitation. The multiplicity of the vertex corrections of the *8* fermion boost lines taken *2* at a time is

$$M = \frac{8!}{2!6!} = 28 \quad . \tag{48}$$

Including the total first order radiative contribution, the geometric magnetic moment term gives

$$\frac{g_1}{2} = \frac{g_0}{2}\left(1 + \frac{28\alpha}{2\pi}\right) = 1.866(1+0.0325) = 1.9267 \tag{49}$$

This result has a *0.5%* discrepancy with the experimental value of the neutron magnetic moment, equal to *1.9130*. We may consider the neutron as a $(p,e,\bar{\nu})$ combination.

# 5. Conclusion.

The geometrical magnetic moment of the proton and the neutron, due to their spins and electromagnetic generators, have a theoretical first order anomalous gyromagnetic Landé *g*-factor which is close to the experimental value with a *0.5%* discrepancy. There may be other corrections in the geometric theory, including necessary higher order terms, which could eliminate the small discrepancy between the theoretical dressed value and experimental values.

# 6. Appendix.

## 6.1 Algebra Representation.

Any element of the geometric algebra [2] may be written in terms of its even and odd parts as

$$a = a_+ + \kappa^0 a_- \tag{50}$$

and may be represented by a matrix of twice dimensions with even components, as follows

$$a \to \begin{bmatrix} a_+ & -\bar{a}_-^\dagger \\ a_- & \bar{a}_+^\dagger \end{bmatrix} . \tag{51}$$

Using this technique we represent the various objects as follows

$$e \to \begin{bmatrix} \eta & -\bar{\xi}^\dagger \\ \xi & \bar{\eta}^\dagger \end{bmatrix} , \tag{52}$$

$$\Gamma \to \begin{bmatrix} \Gamma_+ & -\bar{\Gamma}_-^\dagger \\ \Gamma_- & \bar{\Gamma}_+^\dagger \end{bmatrix} \tag{53}$$

and since

$$\kappa^\mu = \kappa^0 \kappa^{0\dagger} \kappa^\mu = \kappa^0 \bar{\sigma}^\mu , \tag{54}$$

$$\kappa^\mu \to \begin{bmatrix} 0 & -\sigma^\mu \\ \bar{\sigma}^\mu & 0 \end{bmatrix} . \tag{55}$$

## 6.2. Equations of Motion for a Matter Frame.

The explicit expression for the equations of motion of a frame *e* [1] is

$$\kappa^\mu\left(\partial_\mu e - e\Gamma_\mu\right) + \tfrac{1}{2}\kappa^{\hat{\alpha}}\nabla_\nu u_{\hat{\alpha}}^\nu = 0 , \tag{56}$$

which has for even and odd parts,

$$\sigma^\mu\left(\partial_\mu \xi - \xi\Gamma_{+\mu} - \bar{\eta}^\dagger \Gamma_{-\mu}\right) + \tfrac{1}{2}\sigma^{\hat{\alpha}}\nabla_{+\nu}u_{\hat{\alpha}}^\nu \xi = 0 , \tag{57}$$

$$-\bar{\sigma}^\mu\left(\partial_\mu \eta - \eta\Gamma_{+\mu} + \bar{\xi}^\dagger \Gamma_{-\mu}\right) - \tfrac{1}{2}\bar{\sigma}^{\hat{\alpha}}\nabla_{+\nu}u_{\hat{\alpha}}^\nu \eta = 0 , \tag{58}$$

and may be written,

$$\sigma^\mu \nabla_{+\mu} \xi + \tfrac{1}{2} \sigma^{\hat{\alpha}} \nabla_{+\nu} u^\nu_{\hat{\alpha}} \xi = \sigma^\mu \overline{\eta}^\dagger \Gamma_{-\mu} \quad , \tag{59}$$

$$-\overline{\sigma}^\mu \nabla_{+\mu} \eta - \tfrac{1}{2} \overline{\sigma}^{\hat{\alpha}} \nabla_{+\nu} u^\nu_{\hat{\alpha}} \eta = \overline{\sigma}^\mu \overline{\xi}^\dagger \Gamma_{-\mu} \quad . \tag{60}$$

References.